\documentclass[aps,prl,superscriptaddress,twocolumn,longbibliography]{revtex4-1}
\usepackage{dcolumn}
\usepackage{bm}
\usepackage{amssymb}
\usepackage{float}
\usepackage{subcaption}
\usepackage[english]{babel}
\usepackage{tabularx,ragged2e,booktabs}
\usepackage[pdftex]{graphicx}
\usepackage{amsmath,esint}

\usepackage{hyperref}
\usepackage{algpseudocode}

\newcommand{\grad}{\vec{\nabla}}
\newcommand{\tensoriden}{\tensor{\bm{I}}}
\newcommand{\tensorsigma}{\tensor{\bm{\sigma}}}
\newcommand{\tensorsigmanod}{\tensor{\bm{\sigma}_0}}
\usepackage[usenames]{color}

\newcommand{\change}{}

\begin{document}

\widetext

\title{Capillary stress and structural relaxation in moist granular materials}
\author{Tingtao Zhou}
\affiliation{Massachusetts Institute of Technology, Department of Physics}
\author{Katerina Ioannidou}
\affiliation{The MIT / CNRS / Aix-Marseille University Joint Laboratory,
"Multi-Scale Materials Science for Energy and Environment" and Massachusetts Institute of Technology, Department of Civil and Environmental Engineering}
\author{Enrico Masoero}
\affiliation{Newcastle University, School of Engineering, United Kingdom}
\author{Mohammad Mirzadeh}
\affiliation{Massachusetts Institute of Technology, Department of Chemical Engineering}
\author{Roland J.-M. Pellenq}
\affiliation{The MIT / CNRS / Aix-Marseille University Joint Laboratory,
"Multi-Scale Materials Science for Energy and Environment" and Massachusetts Institute of Technology, Department of Civil and Environmental Engineering}
\author{Martin Z. Bazant}
\affiliation{Massachusetts Institute of Technology, Department of Chemical Engineering}\affiliation{Massachusetts Institute of Technology, Department of  Mathematics}

\date{\today}

\begin{abstract}
We propose a theoretical framework to calculate capillary stresses in complex mesoporous materials, such as moist sand, nanoporous hydrates, and drying colloidal films.  Molecular simulations are mapped onto a phase-field model of the liquid-vapor mixture, whose inhomogeneous stress tensor is integrated over Voronoi polyhedra in order to calculate equal and opposite forces between each pair of neighboring grains.  The method is illustrated by simulations of moisture-induced forces in small clusters and \change{heterogeneous porous} packings of spherical grains  using lattice-gas Density Functional Theory.  For a nano-granular model of cement hydrates, this approach reproduces the hysteretic water sorption/desorption isotherms and predicts drying shrinkage strain isotherm in good agreement with experiments.  We show that capillary stress is an effective mechanism for internal stress relaxation in colloidal \change{heterogeneous porous} packings, which contributes to the extraordinary durability of cement paste.
\end{abstract}

\maketitle

Capillary condensation is a ubiquitous process of vapor-liquid phase transition in porous media, such as sand piles, plaster, paints, silica gels, cementitious materials, which has an important yet poorly understood effect on mechanical behavior.  The confined fluid can generate enormous local stresses, as observed in granular material aging\citep{bocquet1998moisture}, wet floor friction \citep{li2007friction}, nano-tribology \citep{binggeli1994influence}, multi-phase immiscible flows\citep{burdine1953relative, corey1954interrelation}, 
 cement drying shrinkage\citep{feldman1968model, igarashi2000autogenous} and in everyday life experiences such as hardening of a drying sponge or building a sand castle on the beach\citep{coussy2011mechanics}. Capillary condensation and evaporation potentially bring undesirable fracture processes, as in drying cracking of colloidal films~\citep{singh2007cracking,routh2013drying} and paints~\citep{tirumkudulu2005cracking}, but capillary stress can also can be exploited in nano-materials fabrication by capillary force lithography~\citep{suh2002capillary}, capillary rise infiltration~\cite{hor2017nanoporous,manohar2017solvent}, evaporation-driven assembly and self-organization~\citep{manoharan2003dense, schnall2006self, lauga2004evaporation, cho2005self}  and composite imbibition~\citep{larson2018fluidflow} and even used to evaluate the atmosphere of planets\citep{fanale1979mars}. 
 
Despite the broad importance of capillary forces, they remain challenging to predict in complex porous materials over the full range of liquid saturation, either in equilibrium or during a dynamical process of drainage/imbibition.
For granular or colloidal materials, existing models addressing partial saturations are oversimplified and only apply either to low humidity (so called pendular/funicular regimes)\citep{scholtes2009capillary, wang2015anisotropy, duriez2016stress} or to idealized geometries (slit/cylindrical independent pores or single sphere against a wall)\citep {ravikovitch2006density, jakubov2002adsorption, gor2010adsorption,andrienko2004capillary}. \change{At higher saturations, models based on geometrical analysis of Young-Laplace equation for smaller clusters~\citep{melnikov2016micro, melnikov2015grain} are proposed but restricted to only equilibrium liquid distributions inside monodisperse packings.} Molecular simulations are also difficult to apply, \change{since the porous structure considered on the mesoscale ($\sim 1~\mu m$) are $\sim$10 orders of magnitude larger than the volume of a single molecule ($\sim 1~nm$)}.

In this Article, we present a theoretical framework to compute capillary forces and associated structural relaxation in complex porous media. The method is illustrated through calculations of moisture-induced capillary stress in colloidal systems, based on lattice-gas simulations of adsorbed water~\citep{kierlik2002adsorption}. As an important application, we calculate drying shrinkage of  cement paste \citep{feldman1968sorption, feldman1968model, parrott1973effect}, where the heterogeneous and multi-scale, fully connected porous network induces  significant mechanical irreversibilities during drying-wetting cycles\citep{maruyama2015bimodal}, disqualifying simple models.

\begin{figure*}
\begin{center}
\begin{subfigure}{0.31\linewidth}
\includegraphics[width=0.98\linewidth]{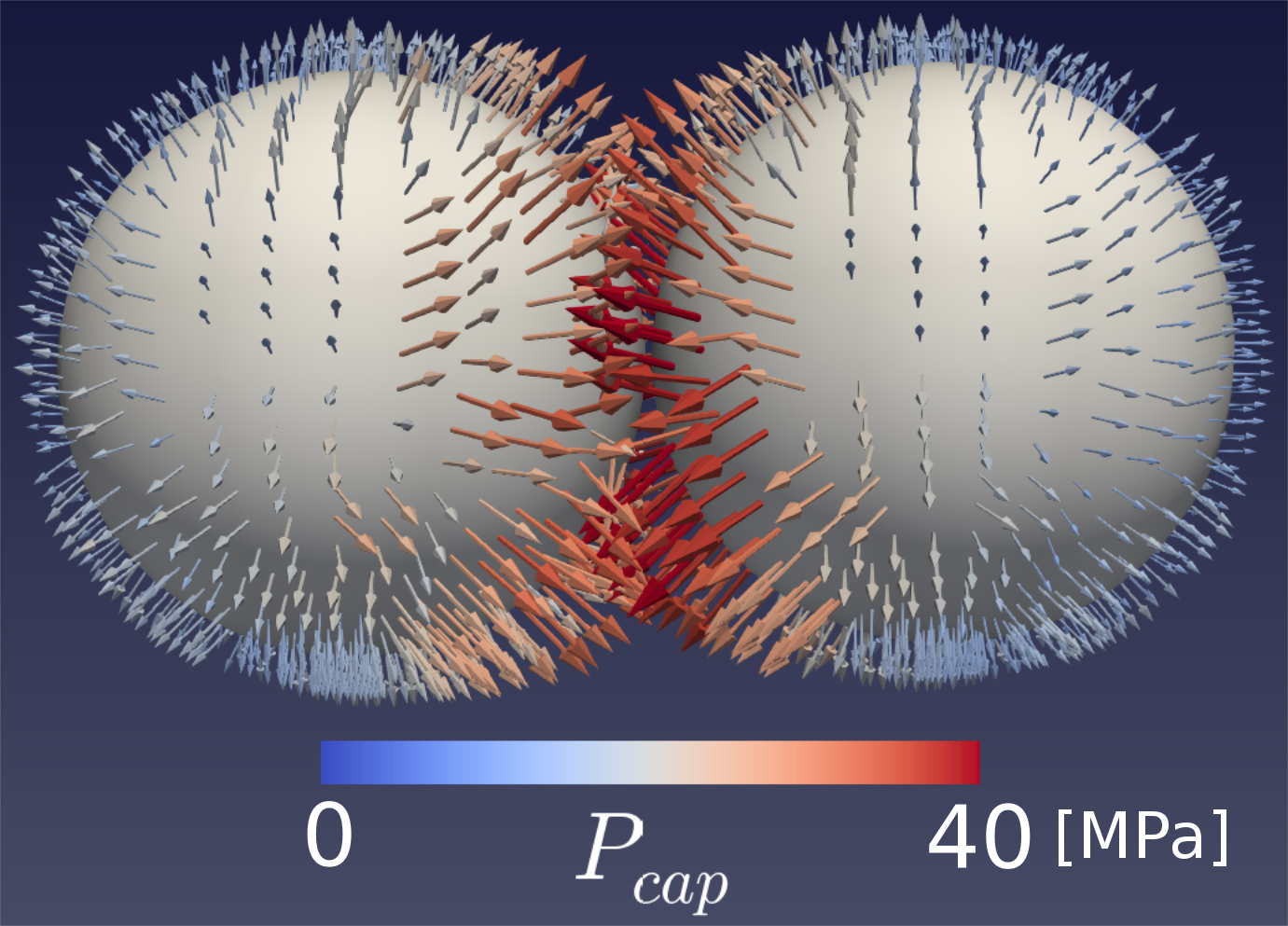}
\subcaption{}
\label{fig:2spheres}
\end{subfigure}
\begin{subfigure}{0.31\linewidth}
\includegraphics[width=0.98\linewidth]{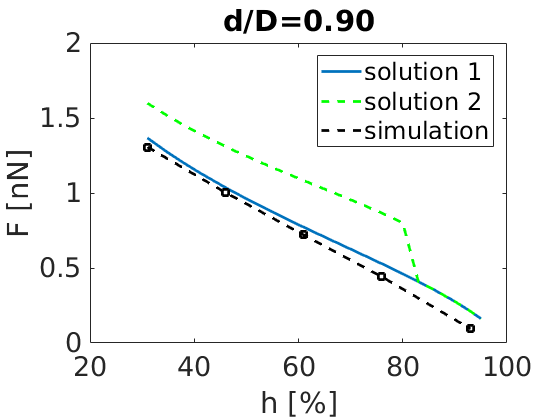}
\subcaption{}
\label{fig:2spheres-force}
\end{subfigure}
\begin{subfigure}{0.31\linewidth}
\centering
\includegraphics[width=0.75\linewidth]{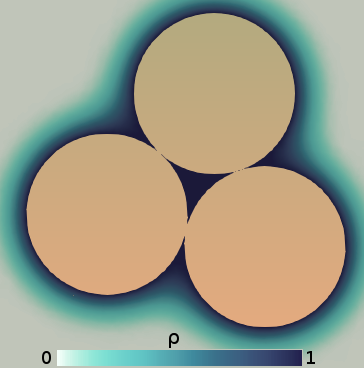}
\subcaption{}
\label{fig:3spheres}
\end{subfigure}
\end{center}
\caption{ (a) Simulation of local stresses in the capillary bridge between two spherical colloidal particles of radius $R=3$~nm at $h=30\%$, and (b) the total capillary force versus humidity, compared with the two analytical solutions of Kelvin-Laplace equation (due to the 3D geometry, see SI Eqn.~17) for sphere-center separation $d=0.9~D$ relative to the sphere diameter, including the surface adsorbed layer. (c) Cross-sectional view of the liquid density $\rho$ filling the region between three touching spherical particles, simulated at $h=35\%$. }
\end{figure*}

{\it Theory---}
For any molecular simulation model of a two phase, liquid-vapor system, the first step of our method is to parameterize the mean-field approximation of a Cahn-Hilliard-like phase-field free energy functional~\citep{cahn1958free,bazant2013theory}
\begin{equation}
\begin{split}
F[\rho] = & \int_V dV \left( f(\rho) + \frac{\kappa}{2}(\grad\rho)^2 + ... \right) \\
& + \iint_{\partial V} d\vec{S}\cdot\left( g(\rho)\vec{n} + h(\rho)\grad\rho + ...\right)
\end{split}
\end{equation}
\change{where the order parameter $0<\rho<1$ here is the normalized liquid density.} The homogeneous free energy density $f(\rho)$, e.g. a regular solution model with particle-vacancy interaction energy $\Omega \rho(1-\rho)$)\citep{cahn1958free, bazant2013theory}, can be fitted to simulation results in periodic cells, and the gradient penalty is related to the liquid/vapor surface tension
\begin{displaymath}
\gamma= \int_{vapor}^{liquid} \kappa (\grad\rho)^2 d\vec{r}\cdot\vec{n}
\end{displaymath}
The planar liquid-vapor interface width for a regular solution model is given by $\lambda=(\kappa/\Omega)^{1/2}$, with interfacial energy $\gamma=(\kappa\Omega)^{1/2}$\citep{cahn1958free, bazant2013theory}.
Solid/fluid surface tension can also be parameterized from simulations, e.g. by interpolating the surface energy term,  represented in the above equation by the surface integral, between stable end states of liquid or vapor.

In some cases, the parameters in the free energy functional can also be derived analytically from the molecular model. For example, for mean-field lattice gas DFT\citep{kierlik2002adsorption} in the continuum limit, we have
\begin{displaymath}
\begin{split}
G & = \int \left(
k_BT \left[ \rho \ln (\rho) + (1-\rho) \ln (1-\rho) \right] -\mu\rho \right) dv   \\
& + \int \left[ \frac{w_{ff}\nu}{4}a^2  (\vec{\nabla} \rho)^2  - \frac{w_{ff}}{2}\nu\rho^2 \right] dv \\
& + \iint_{\partial V} d\vec{S}\cdot \left( w_{sf}\rho\vec{n} - \frac{w_{ff}\nu}{4}a^2\rho\grad\rho \right)
\end{split}
\label{eqn:cahn-hilliard-lattice-gas}
\end{displaymath}
where $w_{ff}$ and $w_{sf}$ are the liquid-liquid and solid-liquid nearest neighbor interactions, $\mu$ the chemical potential for liqdui-vapor equilibirum, $\nu$ the neighbor number and $a$ the size of lattice constant in the lattice gas model. $\partial V$ represents the liquid-solid boundary.

Once we have a suitable free energy, we define the capillary stress tensor, first derived by Korteweg\citep{korteweg1901forme,anderson1998diffuse},
\begin{equation}
\tensor{\bm\sigma} =  \left( p_0(\rho) - \frac{\nu a^2 w_{ff}}{4} (\vec{\nabla}\rho)^2 \right) \tensor{\bm I} + \frac{\nu a^2}{2}w_{ff} \vec{\nabla}\rho \otimes \vec{\nabla}\rho + \tensor{\bm\sigma}_0 
\end{equation}
and use it to express the stress in terms of only the density profile from molecular simulations in the porous structure. (See Supporting Information.)  In principle, surface forces can be calculated by integrating the normal stress over the solid pore walls, but we find that this procedure leads to large errors for complex geometries. 

We thus introduce the second step of our method, which is to utilize the Stokes' theorem\citep{landau1987course} and deform the contour away from the solid surface and integrate the normal stress over a space-filling tessellation of the microstructure. In this way, equal and opposite forces are applied at each face of the tessellation, perfectly satisfying Newton's third law in the fluid, despite fluctuations in the molecular simulation and complex surface geometry.
For colloidal  or granular systems of spherical particles, the most natural choice is the Voronoi tessellation, for which fast algorithms are available, such as the package Voro++\citep{rycroft2009voro++} used below.

Let us apply the method first to the simplest case of two nearly touching grains, which form a stable capillary bridge, using the lattice-gas DFT model for water, as shown in Fig.\ref{fig:2spheres}. Analytical results are available to describe the capillary force\citep{scholtes2009capillary, andrienko2004capillary, cheng2012spontaneous}, and most models for wet granular materials rely on this picture of a capillary brige\citep{kohonen2004capillary,scheel2008liquid, mitarai2006wet, richefeu2006shear, herminghaus2005dynamics, fournier2005mechanical}. The simulated capillary force versus humidity is shown in Fig.\ref{fig:2spheres-force},  in agreement with the solution to the Kelvin-Laplace equation\citep{barrett1951determination,thomson1872kelvinequation, stifter2000theoretical}, assuming bulk water surface tension. The forces are simulated on the adsorption branch and thus compared with the smaller Kelvin radius ``solution~1" \change{(see SI)}, augmented by a wetting layer of thickness 0.25~nm.

To illustrate the challenges posed by any other geometry, consider the next simplest case of 3 spheres nearly in contact as shown in Fig.\ref{fig:3spheres}. At low humidity, the capillary bridge theory  still holds and accurately predicts simulation results, but at high humidity, these bridges coalesce to fill the region between the particles and drastically alters the forces, in a way that only the simulation can easily capture, in part since the liquid-vapor interface takes a non-convex shape in three dimensions. 

\begin{figure}
\includegraphics[width=0.44\linewidth]{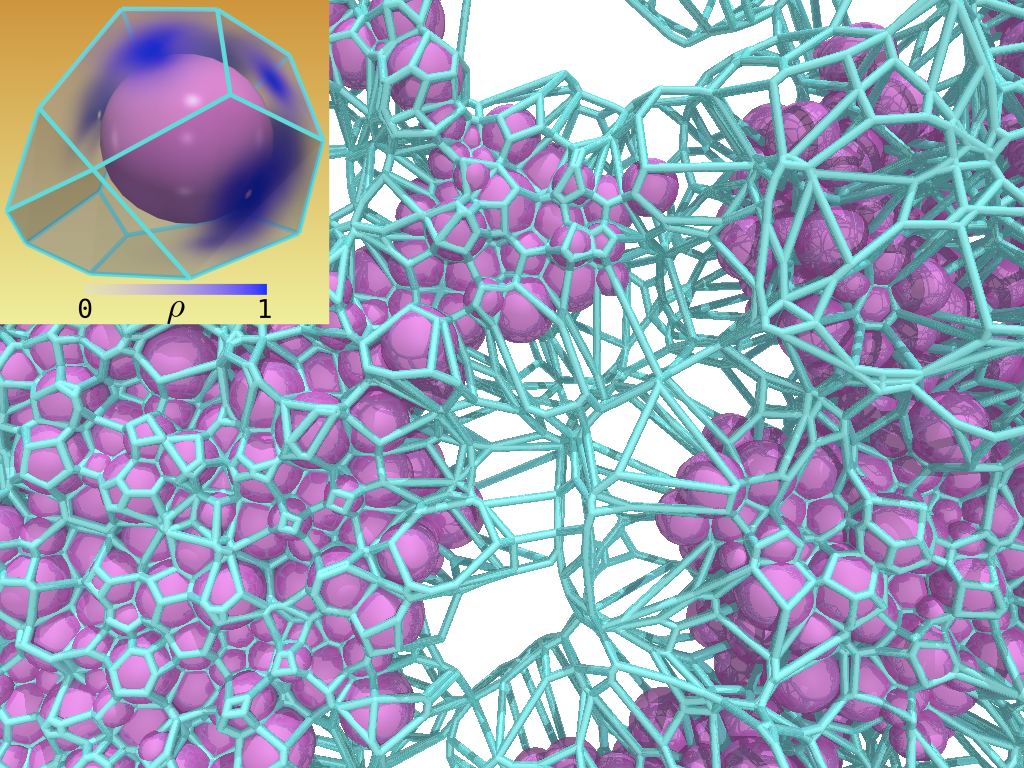}
\caption{Slice of a polydisperse \change{heterogeneous} sphere packing, showing the Voronoi tessellation used to calculate capillary stresses, for a colloidal model of cement paste~\citep{ioannidou2016mesoscale}. The inset shows adsorbed fluid density profile (blue) on the faces of the Voronoi cell for a single grain.}
\label{fig:tessellation}
\end{figure}

Next we consider a 3D \change{heterogeneous porous} packing of colloidal or granular spheres of different sizes. Voro++\citep{rycroft2009voro++} easily computes the tessellation and stresses can be efficiently and accurately computed. Fig.\ref{fig:tessellation} shows the Voronoi tesselation cells around the colloidal nano-grains. Zooming in on a single grain, the inset of Fig.\ref{fig:tessellation} shows the fluid density on the faces of its polygonal Voronoi cell at $h=30\%$. 
Due to the complex granular structure, straightforward integration of the mean stress field over the surface leads to errors that accumulate into a non-zero net force on the solid, violating Newton's third law, but integration over the tessellation allows equal and opposite forces to be assigned to each pair of neighboring particles.
This example demonstrates the capability to predict the stress and deformation of composite materials, subject to either out-of-equilibrium multi-phase flow or equilibrium capillary condensation during drying and wetting processes.

\begin{figure}
\centering
\begin{subfigure}{0.43\linewidth}
\includegraphics[width=0.98\linewidth]{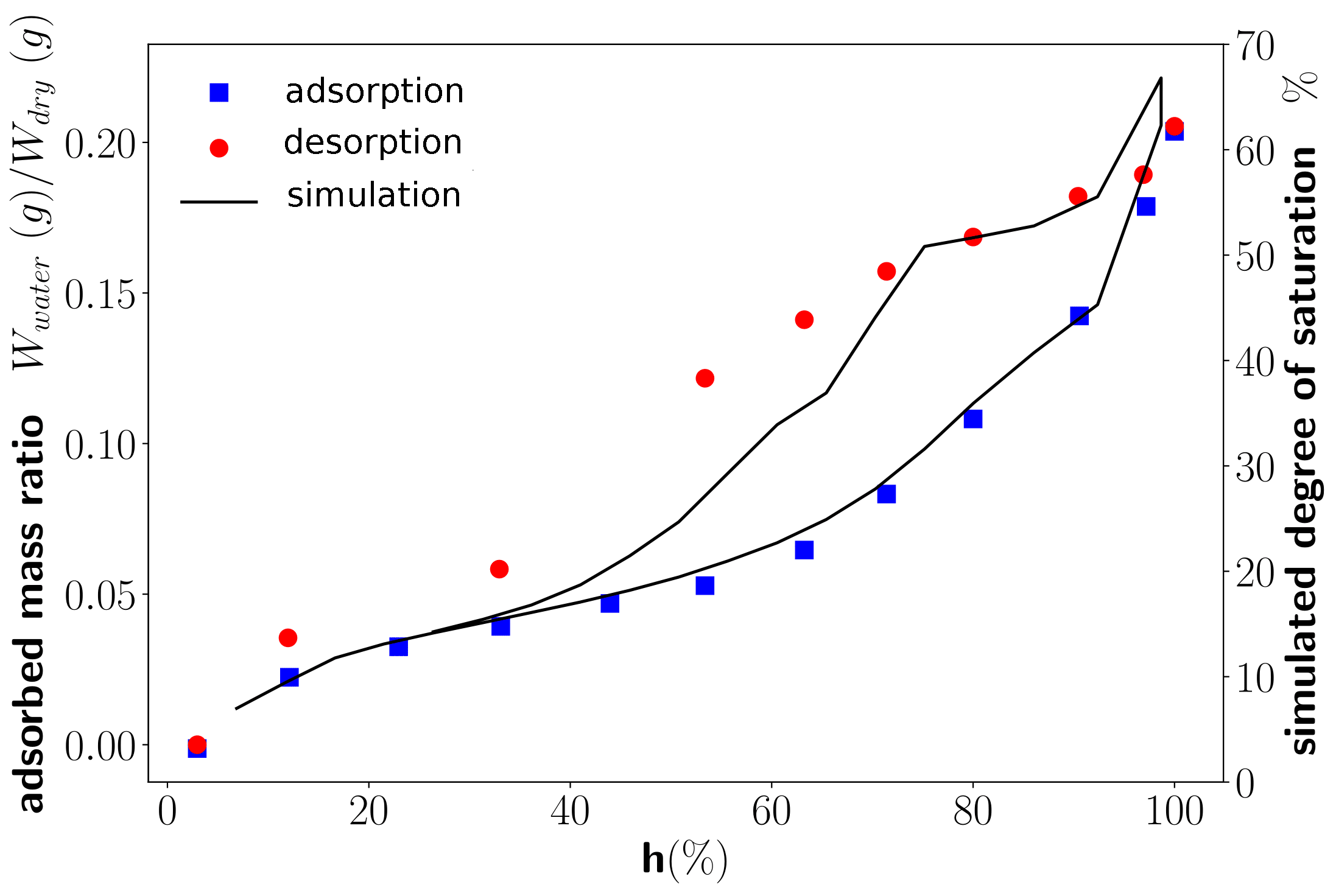}
\subcaption{}
\label{fig:sorption-isotherm}
\end{subfigure}\\
\begin{subfigure}{0.43\linewidth}
\centering
\includegraphics[width=0.98\linewidth]{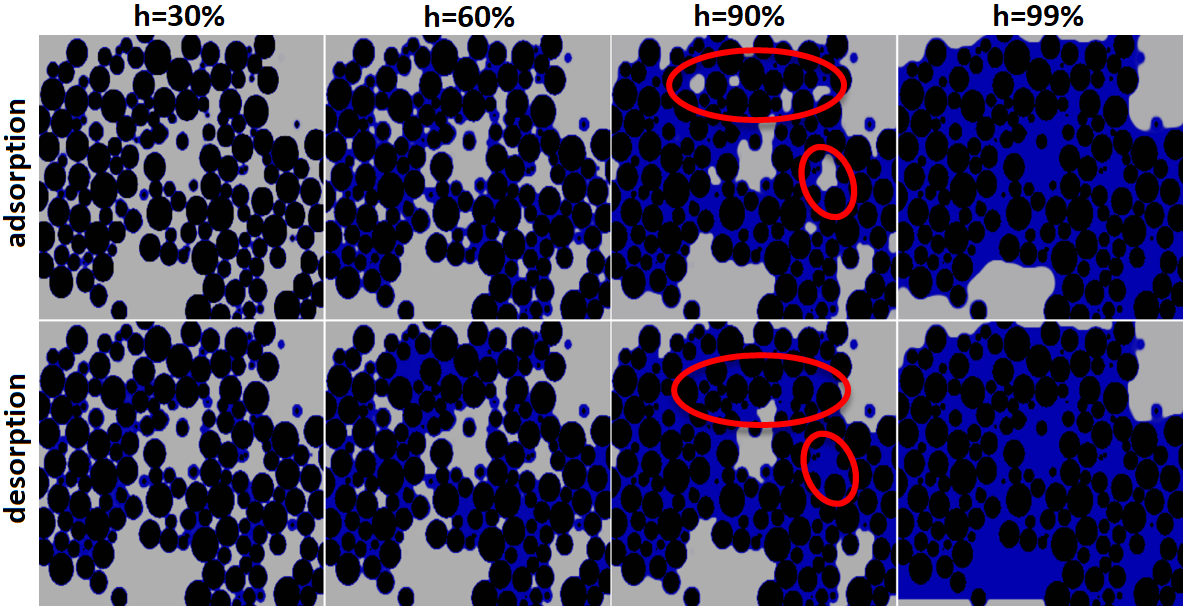}
\subcaption{}
\label{fig:ink-bottleneck}
\end{subfigure}
\caption{ (a) Simulated water adsorption/desorption isotherm of mesoscale cement paste model~\cite{ioannidou2016mesoscale} compared to experimental results~\cite{baroghel2007water}. (b) cross section of the pore network showing water upon adsorption/desorption.  Black---colloidal nano-particles of CSH. Blue---water. Grey---vapor. Ink-bottleneck effect is exemplified by the red-circled areas.}
\label{fig:sorption-results}
\end{figure}

{\it Results---}
As a practical application of our method, we predict capillary stresses and structural relaxation in \change{hardened} cement paste during drying/wetting cycles, which are associated with various degradation mechanisms in concrete structures and pavements.  We simulate capillary condensation in our previously developed colloidal model of cement paste\citep{ioannidou2016mesoscale}, illustrated in Fig.\ref{fig:tessellation}, using interaction parameters imported from atomistic simulations \citep{bonnaud2016interaction,masoumi2017intermolecular}. \change{Within this model the calcium silicate hydrate particles (C-S-H) are approximated as spherical nano-particles and the force field there is designed for hardened cement paste. At very low relative humidities ($h<15\%$) the molecular structure of these nano-particles will also change, where more detailed models on that scale is needed. For our purpose here of studying effects of capillary stress on the pore network, the sphere approximation is sufficient.} First, we show the lattice DFT method\citep{kierlik2002adsorption} for this structure is able to predict the room temperature hysteretic water adsorption/desorption isotherm in agreement with experiments \citep{baroghel2007water} (Fig.\ref{fig:sorption-isotherm}) with no adjustable parameter. \change{To choose a characteristic length scale as the lattice spacing a, we estimate the surface tension which is energy per area:
\begin{equation}
E_{surface}\sim \frac{w_{ff}}{2a^2}
\end{equation}
for nitrogen at $T=77~K$, $E_{surface}\sim8.94mN/m$ which gives $a\sim0.345~nm$; for water at $T=300~K$, $E_{surface}\sim72mN/m$ which gives $a\sim0.24~nm$. Based on the estimates we choose a fine grid cell size of $a=3\AA$.
Another benefit of the very fine lattice spacing  is that it allows us to capture the characteristics of small capillary pores and some gel pores (for categorization of pores see \citep{ioannidou2016mesoscale}) in the system. The high resolution of roughness on the nanoscale also facilitates heterogeneous nucleation on the solid surface. The fluid-fluid interaction $w_{ff}$ is determined by the bulk critical point $k_BT_{cc}=-\frac{\nu w_{ff}}{2}$ where $\nu$ is the number of nearest neighbors on the lattice. The fluid-solid interaction $w_{sf}=2.5~w_{ff}$ is estimated from molecular simulations on isosteric heat during first layer sorption\citep{bonnaud2012thermodynamics} for water in cement paste.
}

Next, we analyze the fluid distribution in the pore network (Fig.\ref{fig:ink-bottleneck}) and calculate the mechanical effect of capillary stress, for the first time to our knowledge in such a complex porous structure.

\begin{figure}
\centering
\begin{subfigure}{0.43\linewidth}
\includegraphics[width=0.98\linewidth]{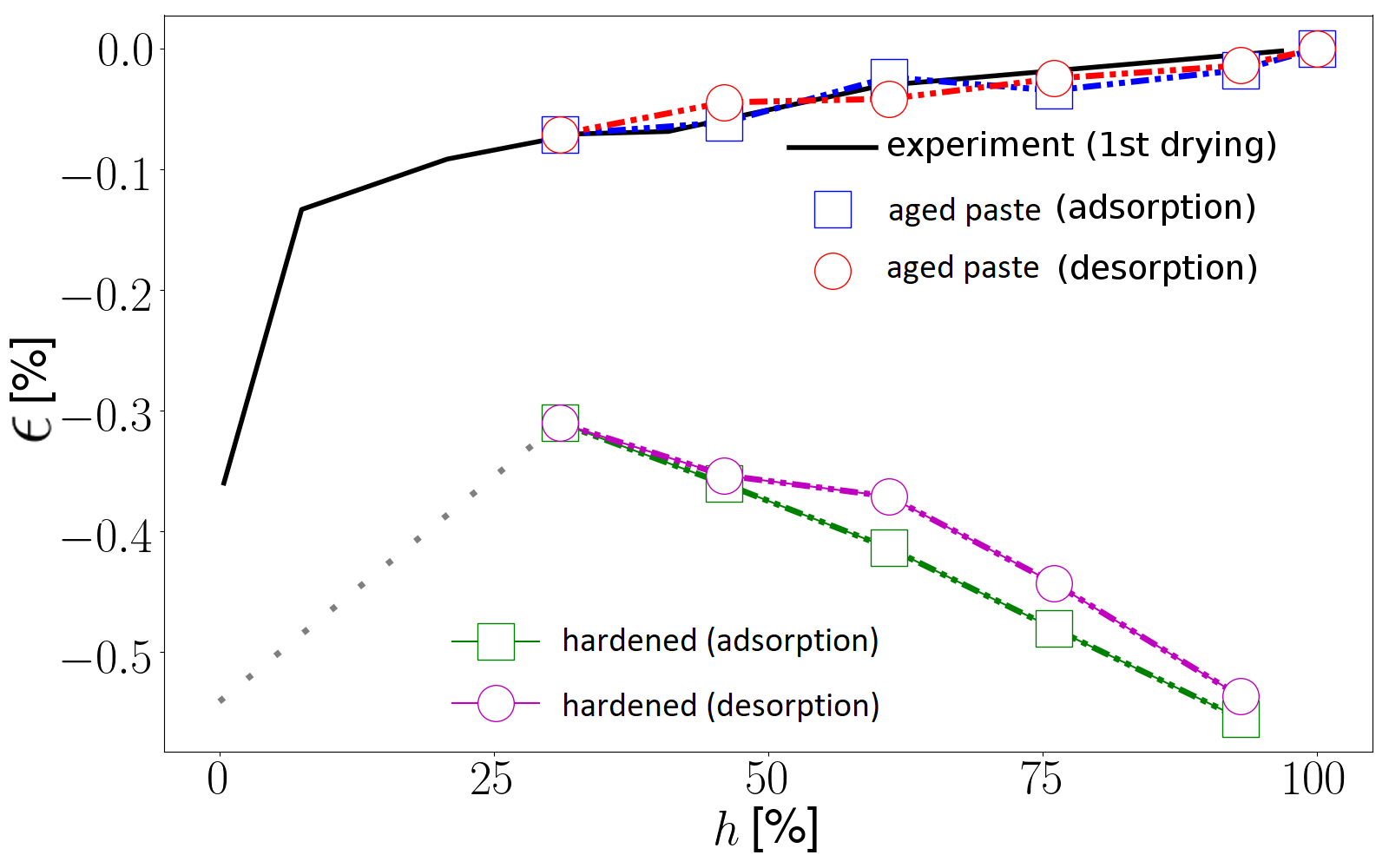}
\subcaption{}
\label{fig:npt-fcap}
\end{subfigure}\\
\centering
\begin{subfigure}{0.43\linewidth}
\includegraphics[width=0.98\linewidth]{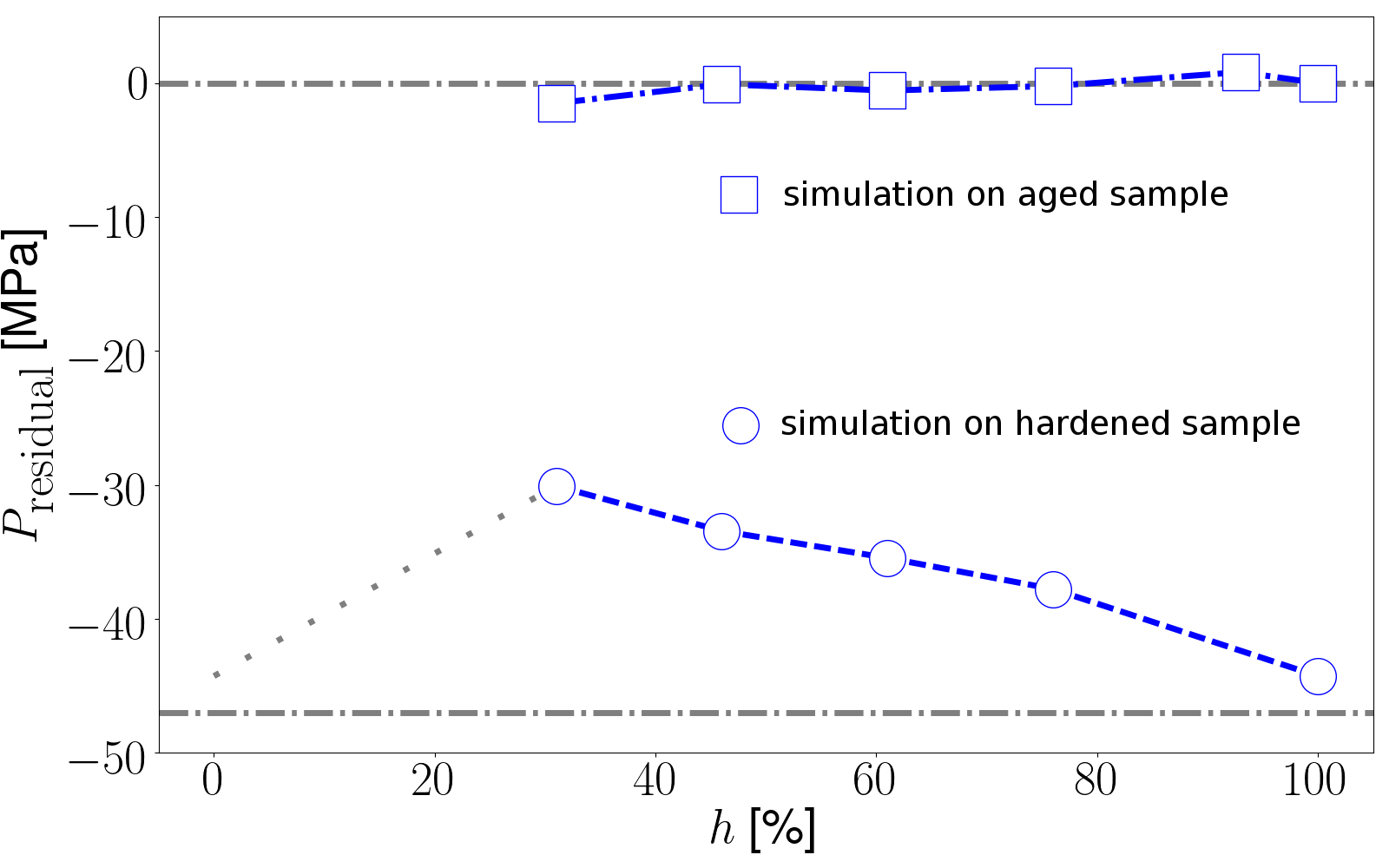}
\subcaption{}
\label{fig:nvt-fcap}
\end{subfigure}
\caption{
\change{
Stress relaxation in two different models of cement paste, a 
\change{hardened} sample with initial tensile eigen-stress $\sim$-47~MPa (lower curves) and a \change{aged} sample, relaxed from the \change{hardened}  to 10~kPa (upper curves). For simulations at constant pressure (NPT) in (a), the \change{aged} sample results are in excellent agreement with experiments on first-cycle drying shrinkage\citep{feldman1968sorption}, and the \change{hardened} sample shrinkage strain indicates stress relaxation assisted by capillary stress. At constant volume (NVT) in (b),  no significant stress produced in the \change{aged} sample, but the \change{hardened} sample's initial solid stress without capillary forces relaxed to -44.27~MPa, blue circles showing more solid stress relaxed by capillary forces. See main text for more discussions.
}
}
\label{fig:npt-nvt-fcap}
\end{figure}

In general, we find that capillary forces facilitate macroscopic stress relaxation in granular  or colloidal cohesive media, which present rough and glassy potential energy landscapes having an abundance of meta-stable states. For the case of cement paste \change{drying shrinkage~\citep{pinson2015hysteresis}}, this mechanism is able to quantitatively explain the experimentally observed volume shrinkage (see Fig.\ref{fig:npt-nvt-fcap}).
Reactive heterogeneous materials such as cement have residual stresses due to the out-of-equilibrium  solidification process\citep{abuhaikal2016expansion,ulm2015shrinkage,abuhaikal2017chatelier}. The cement paste structure analyzed here has accumulated tensile eigen-stress of -47 MPa through the precipitation of nano-grains \citep{ioannidou2014controlling, ioannidou2017inhomogeneity}.  

Reactive heterogeneous materials such as cement contain residual stresses due to the out-of-equilibrium solidification process. \citep{abuhaikal2016expansion,ulm2015shrinkage,abuhaikal2017chatelier} \change{ 
The first cement paste structure analyzed here is such a model that has accumulated tensile eigen-stress of -47 MPa through the precipitation of nano-grains \citep{ioannidou2014controlling, ioannidou2017inhomogeneity}. We refer to this structure as representing a ``hardened'' cement paste, which has GPa modulus. The second structure is by performing molecular dynamics (MD) simulation under 0 external pressure (NPT ensemble) on the hardened model, so that it retains only minimal residual eigen-stress in the solid aggregate ($\sim 1~kPa$), representing an ``aged'' cement paste.
A series of MD simulations are then parallelly carried out using LAMMPS\citep{plimpton1995fast} on the above 2 cement hydrate models:
1) NVT relaxation with reduced temperature $T=0.00015$ corresponding to room temperature. 
2) NPT relaxation at room temperature and 0 ambient pressure. 
In both series of simulations capillary forces calculated at corresponding $h$'s are applied on the nano-grains as force vectors in addition to the original colloidal particle interactions represented by Lennard-Jones potentials.
Simulations were terminated after 500000 MD steps for each $h$ value, with timestep $\delta t=0.0025$ in Lennard-Jones unit, when the system is already converged and stable. In both studies we only focus on the range $30\%<h<100\%$, below which we believe for cement paste other factors come into play and likely take over capillary effects, such as the change in colloidal interactions (due to ion concentration change in the pore solution and surface charge regulation of the CSH nano structure) and interlayer water insertion/evaporation.
}

\change{
Figure \ref{fig:npt-fcap} shows how capillary forces influence the shrinkage strain of the ``hardened'' and ``aged'' models. The aged model exhibits shrinkage  only due to capillary effect whereas the hardened sample shrinks under the combined action of tensile eigen-stresses and capillary forces.  In the aged sample, the "pure capilllary effect" decreases with increasing humidity, and the predicted strain is in excellent agreement with experiment~\citep{feldman1968sorption} for $h>30\%$ on the first drying cycle. In contrast, the hardened sample shrinks more than the aged one and with the opposite trend: the amount of shrinkage strain increases upon increasing humidity: at $h=100\%$ where all pores are filled with water and no capillary forces present, the hardened model experiences length shrinkage strain $\vert\epsilon\vert \sim 0.54\%$; as $h$ decreases, $\vert \epsilon \vert$ monotonically decreases until at $h=31\%$ it displays a minimum shrinkage strain $\vert\epsilon\vert \sim 0.32\%$. The different behaviors observed in the 2 models imply a subtle point: in a non-equilibrium system of colloidal particles (such as the hardened model) capillary forces bias the original potential energy landscape (PEL) so that the system is led to different quenched states. The paths to the new states costs less ``effort'' as measured by the total amount of length change the system undergoes. A related observation is we also notice more significant strain hysteresis in the hardened model, another indication that the synergy of tensile eigen-stress and capillary forces is non-linear and can make the system quench to different quasi-equilibrium states.
}

\change{
In Fig.\ref{fig:nvt-fcap} we investigate stress relaxation under a constant volume constraint. Minimal residual stresses are observed in \change{aged} model for all humidities. Residual stresses in \change{hardened} model are relaxed to -30.52 MPa at $h=31\%$, significantly different than that obtained in fully dry condition  (-44.4 MPa). This conjugate result for hardened model echoes the above NPT shrinkage study, in the sense that the system is able to find paths to less stressful quasi-equilibrium states if its PEL is biased by capillary forces.
}

These results demonstrate that capillary stress at intermediate humidity is an effective mechanism of stress relaxation and therefore beneficial to cement durability.
Although the magnitude of capillary forces on each nano-grain is small ($\sim$1~nN) compared to the grain-grain interactions ($\sim$10~nNs), these forces still have an observable effect on the mesoscale. For materials that are ``softer" than cement paste, capillary forces are expected to have even larger aging effects. Overall capillary forces help to relax eigenstresses without major volume shrinkage. It is remakable to notice the synergetic effect of capillaries and eigen-stresses, reducing shrinkage compared to both dry and fully saturated conditions. \change{We point out that the drying shrinkage of cement paste as a complex porous material involves more factors that should be taken into account, such as the change of the interlayer water and surface charge density \citep{cases1997mechanism, michot2005hydration, rinnert2005hydration, tambach2006hysteresis}, which is most significant at low $h$ values, and here we only focus on the capillary stress effects at intermediate to high $h$.}


In conclusion, we have described a general framework to calculate capillary stress in complex porous media, without any assumptions on the pore morphology or topology, across the full range of liquid saturation. The method is able to predict capillary stress and structural relaxation in cement paste, which has both theoretical and practical importance due to its multi-scale  porous structure and wide usage in everyday life. Other possible applications include the drying and collapse of wet sand castles\citep{pacheco2012sculpting, fall2014sliding}, capillary-driven liquid/solid composite manufacturing\citep{suh2002capillary,larson2018fluidflow}, and the evaporation-driven assembly and self-organization of colloidal particles~\citep{manoharan2003dense, schnall2006self, lauga2004evaporation, cho2005self}. 

\vspace{0.1in}

This work was supported by the Concrete Sustainability Hub at MIT within the framework of ICoME2 Labex Project ANR-11-LABX-0053 and A*MIDEX Project ANR-11-IDEX-0001-02 cofunded by the French program Investissements d'Avenir, managed by the French National Research Agency. The authors thank S. Yip and N. Nadkarni for useful discussions and C. Rycroft for help with visualization. 

\appendix
\section{Appendix}
\subsection{Simulation of adsorption/desorption isotherms}

To simulate the 3D distribution of water inside the porous network, we setup a lattice gas model\citep{kierlik2001capillary, kierlik2002adsorption,coasne2013adsorption} based on a solid distribution of Ioannidou's \textit{et al.} mesoscale C-S-H model \cite{Ioannidou2017jnm}. These models capture structural and mechanical heterogeneities, emerging from the precipitation of C-S-H grains in confined space, and they reproduce a number of experimental properties such as pore size distributions and elastic moduli \cite{ioannidou2014SM,ioannidou2016mesoscale}.
\begin{equation}
E =  -\sum_{<i,j>}w_{ff}\rho_i\rho_j\eta_i\eta_j 
 - \sum_{<i,j>}w_{sf}\left[\rho_i\eta_i(1-\eta_j)+
\rho_j\eta_j(1-\eta_i)\right]
\label{eqn:hamiltonian}
\end{equation}
Where $\rho_i$ denotes the normalized density of fluid on site i, continuously varying from 0 to 1, $\eta_i=0$ or $\eta_i=1$ indicating site i is occupied by solid or not. $w_{ff}$ and $w_{sf}$ are the fluid-fluid interaction and fluid-solid interaction respectively.
This method has its theoretical roots in the family of 3D Ising models\citep{lee1952statistical, young1997spin,fishman1979random,doria2015random,belanger1991random}, successfully applied to Vycor, controlled porous silica glasses and aerogel \citep{detcheverry2004mechanisms, kierlik2001equilibrium}.  To directly compare with sorption experiments, one shall link the chemical potential $\mu$ to relative saturating pressure or relative humidity in the case of water (denoted by the letter h below). In experiments, h is usually controlled by contact with a reservoir of vapor. One can solve the bulk liqiud-vapor system (Equation \ref{eqn:hamiltonian} with no solid, $\eta_i=0$ for any site i) and describe the bulk condensation at $\mu_{sat}=cw_{ff}/2$, which is achieved when $h=100\%$. Then using the relation for ideal gas $\mu-\mu_{sat}=k_BT\ln(h)$ one can get the chemical potential value for a desired h. In the numerical algorithm we use a uniform convergence criteria
\begin{equation}
\delta = \min_i \{\rho_i^{(n+1)}-\rho_i^{n}\}<10^{-6}
\label{eqn:convergence}
\end{equation}
which is usually more strict than that adopted in \citep{kierlik2002adsorption}, quoted here as 
\begin{equation}
\Delta = \frac{1}{N} \sum_i \left( \rho_i^{n+1}-\rho_i^n \right)^2 < 10^{-14}
\label{eqn:convergence-kirlik}
\end{equation}
when simulating complex porous structures. (we present an example comparison in table\ref{tbl:convergence}) In the simulations we choose chemical potential incremental steps corresponding to $\Delta h=3\%$ after testing with smaller steps without significant difference in the sorption results. Our tests of local disorder on the solid surface, such as spatially randomized time-invariant solid-liquid interactions, or adding random solid sites on the original surface to represent molecular disorder, do not show much influence on the sorption results.

\begin{table}[h]
\begin{center}
\setlength{\tabcolsep}{5pt}
\begin{tabular}{| c | c | c | c | c |}
\hline
\hline
$\mu/k_B$ & $\Omega$ & $\delta$ & $\Delta$ & $N_{iter}$ \\ \hline
-2467 &	-7.23E10 &	0.003033 &	4.16E-15  &	65 \\ \hline
-2096 &	-11.55E10 &	0.006889 &	4.27E-15  &	200 \\ \hline
-1935 &	-13.82E10 &	0.004379 &	4.80E-15 &	171 \\ \hline
-1831 &	-15.47E10 &	0.006164 &	4.36E-15 &	130 \\ \hline
-1754 &	-16.86E10 &	0.001608 &	1.03E-15 &	188 \\ \hline
-1692 &	-18.15E10 &	0.006197 &	4.49E-15 &	228 \\ \hline
\hline
\end{tabular}
\caption{comparison between the convergence criteria: the error $\delta$ corresponds to Eqn.\ref{eqn:convergence}; the error $\Delta$ is from \citep{kierlik2002adsorption} (see Eqn.\ref{eqn:convergence-kirlik}). The simulation was considered as converged when $\Delta < 10^{-14}$, but still not converged according to Eqn.\ref{eqn:convergence}. $\mu$ is the chemical potential, $\Omega$ the total grand potential of the system. The last column shows the number of iterations $N_{iter}$ when convergence according to $\Delta<10^{-14}$ reached. (During our simulation on the mesoscale porous model of cement we limit the maximum number of iterations for 1 single relative humidity to 600) Here none of the $\mu~s$ are considered converged by the criteria of Eqn.\ref{eqn:convergence}}
\label{tbl:convergence}
\end{center}
\end{table}

As we are interested in a quantitative comparison with experimental sorption results, we need to define a physically meaningful length scale determined by the lattice spacing size. The parameters $w_{ff}$ and $w_{sf}$ are in units of energy. To arrive at a characteristic length scale as the lattice spacing a, we estimate the surface tension which is energy per area:
\begin{equation}
E_{surface}\sim \frac{w_{ff}}{2a^2}
\end{equation}
for nitrogen at $T=77~K$, $E_{surface}\sim8.94mN/m$ which gives $a\sim0.345~nm$; for water at $T=300~K$, $E_{surface}\sim72mN/m$ which gives $a\sim0.24~nm$. Based on the estimates we choose a fine grid cell size of $a=3\AA$.
Another benefit of the very fine lattice spacing  is that it allows us to capture the characteristics of small capillary pores and some gel pores (for categorization of pores see \citep{ioannidou2016mesoscale}) in the system. The high resolution of roughness on the nanoscale also facilitates heterogeneous nucleation on the solid surface. The fluid-fluid interaction $w_{ff}$ is determined by the bulk critical point $k_BT_{cc}=-\frac{cw_{ff}}{2}$ where c is the number of nearest neighbors on the lattice. The fluid-solid interaction $w_{sf}=2.5~w_{ff}$ is estimated from molecular simulations on isosteric heat during first layer sorption\citep{bonnaud2012thermodynamics} for water in cement paste.

\subsection{Phase-field formulation and capillary stress calculation}

The lattice model can be viewed mathematically in the continuum limit a phase-field model. The free energy can be written as
\begin{widetext}
\begin{equation}
g({\rho}) = \iiint\limits_V dV \left\lbrace -\frac{c}{2} w_{ff}\rho^2 + \frac{ca^2}{4} w_{ff}(\grad\rho)^2 - \mu\rho + k_BT \left[\rho \ln\rho +
(1-\rho)\ln(1-\rho)\right] \right\rbrace 
 + \oiint\limits_{\partial V} d\vec{S}\cdot \left( w_{sf}\rho \vec{n}_{sf} - \frac{ca^2}{4} w_{ff} \rho\grad\rho \right)
\end{equation}
\end{widetext}
where $\vec{n}_{sf}$ is the boundary normal vector at liquid-solid surface denoted by $\partial V$, pointing from liquid outward into solid; $a=0.3~nm$ is the choice of lattice spacing.

Now one minimizes the free energy by taking its variational under the constraint that on the interface $\delta\rho|_{\partial V}=0$. Then the interfacial term yields
\begin{equation}
\begin{split}
\delta g_1 = & \delta \oiint\limits_{\partial V} d\vec{S}\cdot \left( w_{sf}\rho \vec{n}_{sf} - \frac{ca^2}{4} w_{ff} \rho\grad\rho \right)\\
= & \oiint\limits_{\partial V} d\vec{S}\cdot \left( w_{sf}\delta\rho \vec{n}_{sf} - \frac{ca^2}{4} w_{ff} \grad(\rho\delta\rho) \right)\\
= & - \frac{ca^2}{4} w_{ff} \oiint\limits_{\partial V} d\vec{S}\cdot \rho\delta\grad\rho
\end{split}
\end{equation}
The bulk term contributes as 
\begin{widetext}
\begin{equation}
\begin{split}
\delta g_0 = & \iiint\limits_{V} dV \left\lbrace - cw_{ff}\rho\delta\rho + \frac{ca^2}{2} w_{ff}(\grad\rho)\cdot\delta\grad\rho - \mu\delta\rho + k_BT\delta\rho \ln\left(\frac{\rho}{1-\rho}\right) \right\rbrace \\
= & \iiint\limits_{V} dV \left\lbrace - cw_{ff}\rho - \frac{ca^2}{2} w_{ff}(\grad^2\rho) - \mu + k_BT \ln\left(\frac{\rho}{1-\rho}\right) \right\rbrace \delta\rho + \frac{ca^2}{2}w_{ff}\oiint\limits_{\partial V} d\vec{S}\cdot\delta\rho\grad\rho
\end{split}
\end{equation}
\end{widetext}

If we further insist the boundary variational of derivative $\delta\frac{\partial}{\partial \vec{n}}\rho|_{\partial V}=0$ guarantees the last surface term in $\delta g_0$ and $\delta g_1$ are both 0. The bulk term gives the equilibrium condition
\begin{equation}
-cw_{ff}\left(\rho+\frac{a^2}{2}\grad^2\rho\right) + k_BT\ln\left(\frac{\rho}{1-\rho}\right) - \mu = 0
\end{equation}

The general form of the Landau-Ginsberg free energyas a functional of the field $\rho$ and its gradient respecting symmetries is
\begin{equation}
\Omega = \int f(\rho, \grad \rho) = \int f_0(\rho) + {f_1(\rho)}(\grad \rho)^2 + O((\grad \rho)^4)
\label{eqn-phase-field:general-landau-ginsberg-form}
\end{equation}
here we cut off at the quardratic term of the gradient and $f_1(\rho)>0$ so that thermodynamic stability is gauranteed.
To derive the capillary stress tensor field from the free energy density, one notices the thermodynamic relation that the pressure is $p=\mu\rho-f_0(\rho)=\rho f_0'-f_0$ for a homogeneous system. The stress tensor reads as $\tensorsigmanod=\left( \rho f_0'-f_0 \right)\tensoriden$.
At mechanical equilibrium 
\begin{equation}
-\grad \cdot\tensorsigmanod=-\grad p=-\grad (\mu\rho)+\grad f_0=-\rho\grad f_0'=-\vec{F}_{ext}
\end{equation}
hence
\begin{equation}
-\grad \cdot(\tensorsigmanod-\mu\rho\tensoriden) = \grad (\rho f_0') - \rho\grad f_0' = \grad\cdot (f_0\tensoriden) = \frac{\partial f_0}{\partial \rho} \grad \rho
\label{eqn-stress:generalize-starting-point}
\end{equation}

For inhomogeneous system where $f=f(\rho, \grad \rho)$, 
one way to generalize the pressure to a stress tensor is starting with Eqn.\ref{eqn-stress:generalize-starting-point}  so that now 
\begin{equation}
\grad \cdot(\tensorsigma-\mu\rho\tensoriden)=-\frac{\delta f(\rho, \grad \rho)}{\delta \rho} \grad \rho
\end{equation}
where higher order derivatives of $\rho$ is neglected.
The second order term in Eqn.\ref{eqn-phase-field:general-landau-ginsberg-form} involving $\grad\rho$ is written as $f_1=f_1(\rho)(\grad \rho)^2 = \left(\grad g_1(\rho)\right)^2$ where $g_1(\rho)=\int \sqrt{f_1(\rho)}d\rho$.
Using Stoke's theorem
\begin{equation}
\begin{split}
\int f_1 = & \int \left(\grad g_1(\rho)\right)^2 = \int \grad\cdot\left(g_1(\rho)\grad g_1(\rho)\right)-g_1(\rho)\grad^2g_1(\rho)\\
= & -\int g_1(\rho)\grad^2g_1(\rho)
\end{split}
\end{equation}
Now $f(\rho, \grad \rho)=f_0+f_1^*$ where $f_1^*=-g_1(\rho)\grad^2g_1(\rho)$

Because of the identities
\begin{equation}
\begin{split}
\vec{u}\times(\grad \times\vec{u}) & = \frac{1}{2}\grad (\vec{u}\cdot\vec{u})^2 - \vec{u}\cdot\grad \vec{u}\\
\vec{u}\cdot\grad \vec{u} & = \grad \cdot(\vec{u}\otimes\vec{u}) - \vec{u}\grad \cdot\vec{u} \\
\grad \left(\vec{u}\cdot\vec{u}\right) & = \grad \cdot \left( \vec{u}\cdot\vec{u} \tensoriden \right)
\end{split}
\end{equation}
and notice $\grad\times\grad g_1(\rho)=0$ we then have
\begin{equation}
\begin{split}
\frac{\delta f_1^*}{\delta \rho}\grad\rho & = -2\grad g_1(\rho)\grad^2g_1(\rho)
\\
& = -2 \left[ \grad\cdot\left( \grad g_1(\rho)\otimes\grad g_1(\rho) \right) - \grad g_1(\rho)\cdot\grad \grad g_1(\rho) \right] 
\\
& = -2 \grad\cdot\left( \grad g_1(\rho)\otimes\grad g_1(\rho) \right) + \left( \grad g_1(\rho)\cdot\grad g_1(\rho) \right)^2\tensoriden
\end{split}
\label{eqn-stress:core-derivative}
\end{equation}

Finally we arrive at the generalized inhomogeneous stress tensor
\begin{equation}
\begin{split}
\tensorsigma = 2 \grad g_1(\rho)\otimes\grad g_1(\rho) + \left( \mu\rho - (\grad g_1(\rho))^2 \right)\tensoriden
\end{split}
\end{equation}

After one gets the capillary stress as a 3D field, it is straightforward to integrate this field over surfaces of nano-grains to yield the forces on them. One demonstration is shown in Fig.~1(a) of the main text where the capillary attraction between 2 identical spherical solid of radii R with distance between centers D is calculated. Analytical solutions based on Kelvin-Laplace equation can be constructed by 
\begin{equation}
\begin{split}
\frac{k_BT \ln(h)}{\gamma V_m} & = \frac{1}{r_1} - \frac{1}{r_2} \\
(r_1+R)^2 & = (r_1+r_2)^2 + (\frac{d}{2})^2
\end{split}
\end{equation}
which is effectively a set of quadratic equation that generally gives rise to at least 2 solutions. Starting from dry condition on the adsorption branch the equilibrium state corresponds to the solution with smaller capillary bridge.

\subsection{Relaxation with capillary stress}

A series of MD simulations were carried out using LAMMPS\citep{plimpton1995fast} on the cement hydrate model (hereafter refered as M0). 1) NVT relaxation with reduced temperature $T=0.00015$ corresponding to room temperature, hereafter refered as M1. 2) NVT relaxation same as 1) but with capillary forces applied on the grains during the entire simulation, keeping them as constant force vectors, refered as MF1. 3) NPT relaxation at room temperature and 0 ambient pressure, refered as MP1. 4) NPT relaxation same as 3) but with capillary forces constantly applied during entire simulation, refered as MPf1. All simulations were terminated after 500000 MD steps with timestep $\delta t=0.0025$ in Lennard-Jones unit when the system is already converged and stable.
The original model (M0) retains eigen-stress corresponding to a pressure of -47 MPa (minus sign denoting tensile stress). This was mildly relaxed to -44.4 MPa in M1. MF1 at relative humidity $h=31\%$ relaxes the pressure to -30.52 MPa, showing capillarity as an effective mechanism of stress relaxation. MP1 shrinks the simulation box length by $0.54\%$, while MPF1 at $h=31\%$ generates bulk length shrinkage of $0.32\%$. The overall shrinkage strain calculated in both ways are in agreement with the experimental measurements\citep{mills1966effects, maruyama2015bimodal}.

\bibliographystyle{ieeetr}

\end{document}